\pdfoutput=1

\documentclass[showpacs,nofootinbib,showkeys,eqsecnum,prd,aps,preprint]{revtex4-2}

\usepackage[english]{babel}

\usepackage{amssymb}
\usepackage{amsmath}
\usepackage{amsfonts}
\usepackage{latexsym}
\usepackage{mathrsfs}
\usepackage{bm}
\usepackage{tensor}
\usepackage{hyperref}
\usepackage{graphicx}

\def\Sp{\mathop{\rm Sp}\nolimits\,}

\def\dac{\displaystyle\frac}
\def\dil{\displaystyle\int\limits}

\usepackage{diagbox}

\def\pa{\partial}

\def\BI{{\mathbb I}}
\def\CP{{\mathcal P}}

\def\CL{{\mathcal L}}

\def\BR{{\mathbb R}}

\def\BFC{{\bf C}}

\def\Sp{\mathop{\rm Sp}\nolimits\,}

\def\diag{\mathop{\rm diag}\nolimits\,}

\def\dac{\displaystyle\frac}

\def\dil{\displaystyle\int\limits}
\def\{{\lbrace}
\def\}{\rbrace}

\def\Or{{\rm O}}

\begin{document}

\title{Family of asymptotic solutions to the two-dimensional kinetic equation with a nonlocal cubic nonlinearity}

\author{Alexander V. Shapovalov}
\email{shpv@phys.tsu.ru}
\affiliation{Department of Theoretical Physics, Tomsk State University, Novosobornaya Sq. 1, 634050 Tomsk, Russia}
\affiliation{Laboratory for Theoretical Cosmology, International Centre of Gravity and Cosmos, Tomsk State University of Control Systems and Radioelectronics, 40 Lenina av., 634050 Tomsk, Russia}

\author{Anton E. Kulagin}
\email{aek8@tpu.ru}
\affiliation{Division for Electronic Engineering, Tomsk Polytechnic University, 30 Lenina av., 634050 Tomsk, Russia}
\affiliation{Laboratory of Quantum Electronics, V.E. Zuev Institute of Atmospheric Optics, SB RAS, 1 Academician Zuev Sq., 634055 Tomsk, Russia}

\author{Sergei A. Siniukov}
\email{ssaykmh@yandex.ru}
\affiliation{Department of Theoretical Physics, Tomsk State University, Novosobornaya Sq. 1, 634050 Tomsk, Russia}

\begin{abstract}
We apply the original semiclassical approach to the kinetic ionization equation with the nonlocal cubic nonlinearity in order to construct the family of its asymptotic solutions. The approach proposed relies on an auxiliary dynamical system of moments of the desired solution to the kinetic equation and the associated linear partial differential equation. The family of asymptotic solutions to the kinetic equation is constructed using the symmetry operators acting on functions concentrated in a neighborhood of a point determined by the dynamical system. Based on these solutions, we introduce the nonlinear superposition principle for the nonlinear kinetic equation. Our formalism based on the Maslov germ method is applied to the Cauchy problem for the specific two-dimensional kinetic equation. The evolution of the ion distribution in the kinetically enhanced metal vapor active medium is obtained as the nonlinear superposition using the numerical--analytical calculations.\\
\end{abstract}



\maketitle

\section{Introduction}
\label{sec:int}
Kinetic equations are the theoretical footing for the dynamic phenomena of various nature that occur in physical systems of many interacting elements (particles). Examples of such system are diluted gases, gas discharges, a~plasma, processes of coagulation~\cite{Mitrophanov20142347}, and~biological systems such as, e.g.,~population systems \protect{\cite{murray2001,Gluzman2018,Aranson2006}}. In some systems, nonlocal collective (averaged) interactions of elements substantially contribute to dynamics. Such interactions are modeled by integral terms in kinetic equations that become integro-differential. In spatially heterogeneous kinetic phenomena, the~interelement interactions occur along with the diffusion. Then, the~model kinetic equation belongs to the class of reaction-diffusion (RD) equations. The~study of RD equations with both local and nonlocal terms have formed an independent branch of mathematical~physics.

Due to the mathematical complexity of the study of RD equation with nonlocal interactions, methods of computer modeling prevail here. However, the~demand for the analytical methods stimulates the development of approximate and asymptotically exact solutions. For a number of RD kinetic equations with nonlocal interactions, one can succeed using the WKB--Maslov theory of semiclassical approximation or the Maslov complex germ method~\cite{Maslov1,Maslov2,BeD2}. Based on the WKB--Maslov theory, the~method of semiclassical asymptotics was developed for a generalized Fisher--Kolmogorov--Petrovskii--Piskunov equation (Fisher--KPP) with a quadratic nonlocal term in~\cite{shap2009,fkppshap18} and for the nonlocal Gross--Pitaevskii equation in~\cite{shapovalov:BTS2,sym2020}.

In this work, using the results of~\cite{fkppshap18,shapkul21}, we construct semiclassical asymptotics for the model kinetic equation with the nonlocal cubic nonlinearity of the form
\begin{equation}
\begin{gathered}
 \pa_t u(\vec{x},t)=D \tilde{D}_a(t) \Delta_x u(\vec{x},t)+a(\vec{x},t) u(\vec{x},t)\\
 - \varkappa u(\vec{x},t) \dil_{{\mathbb{R}}^n}d\vec{y} \dil_{{\mathbb{R}}^n}d\vec{z}\, b(\vec{x},\vec{y},\vec{z},t)u(\vec{y},t) u(\vec{z},t).
 \end{gathered}
 \label{int1}
\end{equation}
Here, $t$ is a time, $u(\vec{x},t)$ is a distribution function (e.g., the~particle density in a system), \mbox{$\pa_t=\dac{\pa}{\pa t}$.} In~a general case, the~ method under consideration is applicable for $n$-dimensional space, $\vec{x}=(x_1,x_2,\ldots,x_n)=(x_i)\in{\mathbb{R}}^n$. The~nonlinearity parameter $\varkappa$ and the small diffusion parameter $D$ are introduced explicitly for the sake of convenience. The~$n$-dimensional Laplace operator in the Cartesian space $\vec{x}\in{\mathbb{R}}^n$ is denoted by $\Delta_x$. The~coefficients $a(\vec{x},t)$ and $b(\vec{x},\vec{y},\vec{z},t)$ are smooth functions of their spatial arguments that grow not faster than polynomially at each point $t$.

In the physical two-dimensional or three-dimensional space, Equation \eqref{int1} is considered as a model of the optical metal vapor active medium (MVAM) excited by an electrical discharge~\cite{shapkul21}. The~MVAM is a mixture of a buffer inert gas and metal vapors in a gas discharge tube (GDT) (see~\cite{little98,sabotinov2007} and references therein). In the active medium excited by an electrical discharge, the~ionization and recombination processes are mainly caused by the inelastic electron impact. For~typical pressures of a buffer gas and metal vapors, preferentially metal atoms are ionized in the mixture. The~process of triple recombination of an ion with two electrons is responsible for the deionization (see, e.g.,~\cite{gurpit64}). Such dense plasma formed by metal ions and electrons can be considered as quasineutral. The~contracted electrical discharge generates ions and electrons localized in the neighborhood of the GDT center. It means that the concentration of the charges rapidly decreases with the distance from the GDT center. In~\cite{kulopt19,shapkul21}, the~description of the plasma kinetics under assumptions made was based on the following equation:
\begin{equation}
\pa_t n_i = D_a(t)\Delta_x n_i + q_i n_e n_{neut} - q_{tr} n_i (n_e)^2,
\label{int2}
\end{equation}
where $\vec{x}$ are Cartesian coordinates of a point in ${\mathbb{R}}^2$ or ${\mathbb{R}}^3$ depending on the problem statement. The~quantity $q_i=q_i(\vec{x},t)$ is the kinetic coefficient for the electron impact ionization of neutral atoms with a concentration $n_{neut}=n_{neut}(\vec{x},t)$ ($n_e=n_e(\vec{x},t)$ is an electron concentration). In~the same sense, the~coefficient $q_{tr}=q_{tr}(\vec{x},t)$ meets the process of triple recombination of ions with a concentration $n_i=n_i(\vec{x},t)$. We assume the plasma to be dense so that the triple recombination dominates over the dielectronic recombination. The~coefficient $D_a(t)$ is an ambipolar diffusion coefficient. The~dependence of $q_i$, $n_{neut}$, $q_{tr}$, and~$D_a$ on $\vec{x}$ and $t$ is due to their dependence on electron temperature that substantially depends on $\vec{x}$ and $t$.

Assuming the quasineutrality of plasma, the~concentrations of ions and electrons are the same, i.e.,
\begin{equation}
n_e(\vec{x},t)=n_i(\vec{x},t).
\label{int3}
\end{equation}
Then, for~given $a(\vec{x},t)=q_i(\vec{x},t)n_{neut}(\vec{x},t)$ and $q_{tr}(\vec{x},t)$, Equation \eqref{int2} becomes closed and determines the concentration $n_i(\vec{x},t)$ for the given initial and boundary~conditions.

To apply the method of semiclassical asymptotics borrowed from papers \cite{fkppshap18,shapkul21}, we write Equation \eqref{int2} in the nonlocal form \eqref{int1}. For~the space ${\mathbb{R}}^2$ or ${\mathbb{R}}^3$, the~function $b(\vec{x},\vec{y},\vec{z},t)$ is the probability density of a triple recombination due to the collision of an ion with two electrons. The~ambipolar diffusion coefficient $D_a(t)$ in Equation \eqref{int2} is written as $D \tilde{D}_a(t)$ where $D$ is the asymptotic small~parameter.

In this work, following the method of semiclassical asymptotics~\cite{shap2009,fkppshap18}, we have constructed approximate solutions of Equation \eqref{int1} in an explicit analytical form for the special set of equation coefficients. The~obtained expressions are leading terms of semiclassical expansion for the solutions of Equation \eqref{int1} within the accuracy of ${\rm O}(D^{3/2})$ in the following class ${\mathcal P}_D^t$ of trajectory concentrated functions (TCF):
\begin{equation}
 {\mathcal P}_D^t=
\biggl\{\!\Phi :\Phi(\vec x,t,D)= \varphi\Bigl(\!\displaystyle\frac{\Delta\vec
x}{\sqrt{D}},t,D\!\Bigr)
\exp\Bigl[\!\displaystyle\frac{1}{D}S(t,D)\Bigr]\!\!\biggr\},
 \label{fkpp4}
 \end{equation}
where $\Phi(\vec x,t,D)$ is a generic element of the class ${\mathcal P}_D^t$; $\Delta \vec{x}=\vec{x}-\vec{X}(t,D)$; the real function $\varphi(\vec{\eta},t,D)$ belongs to the Schwartz space ${\mathbb{S}}$ in variables $\vec{\eta}$, smoothly depends on $t$, and~regularly depends on $\sqrt{D}$ as $D\to 0$. The~real smooth functions $S(t,D)$ and $\vec{X}(t,D)$, characterizing the class ${\mathcal P}_D^t$, regularly depend on $\sqrt{D}$ as $D\to 0$ and are to be~determined.

Note that the approach proposed can be useful for other models based on nonlinear equations similar to \eqref{int2}. Nonlinear kinetic equations arise in various areas such as cosmology models (see review~\cite{Bamba2012155}), superfluidity models~\cite{pitaevskii2016}, etc.

In the next section, we expound the main ideas of the our approach and basic notations, and~we introduce the linear equation associated with the original kinetic equation whose solutions include the asymptotic solutions to the original kinetic equation. In~Section~\ref{sec3}, we obtain the particular solution to the associated linear equation for the special choice of equation coefficients in the two-dimensional case. In~Section~\ref{sec4}, the~main object of Maslov theory, the~germ, is obtained. Here, we present the symmetry operators to the associated linear equation and construct the family of its solutions. In~Section~\ref{sec5}, we apply the algebraic conditions on the solutions to the associated linear equation that allow us to obtain the countable set of new asymptotic solutions to the nonlinear kinetic equation. Moreover, the~new method for constructing asymptotic solutions to the Cauchy problem for the kinetic equation based on the nonlinear superposition principle is proposed. In~Section~\ref{sec6}, the~specific physically motivated example of the two-dimensional kinetic equation is considered. We illustrate the general formalism of our semiclassical approach by constructing the evolution of the initial ion distribution in the relaxing kinetically enhanced active medium. In~Section~\ref{concl}, we conclude with some remarks.

\section{Leading term of semiclassical asymptotics for the Cauchy problem solution}
\label{sec2}
In this section, we recapitulate the general scheme for the method of constructing the leading term of semiclassical asymptotics for Equation \eqref{int1}. The~detailed description of this method can be found in~\cite{shapkul21}.

According to~\cite{shapkul21}, the~following asymptotic estimates hold for functions $u(\vec{x},t,D)$ from the class ${\mathcal P}_D^t$:
\begin{equation}
\dac{1}{||u||}||\hat{p}^k \Delta x^l u||=\Or(D^{(k+1)/2}), \qquad \dac{1}{||u||}\Big\Vert\hat{T}\big(\vec{X}(t,D),t\big)u\Big\Vert=\Or(D),
\label{est1}
\end{equation}
where $\Delta\vec{x}=\vec{x}-\vec{X}(t,D)$, $\hat{\vec{p}}=D\nabla$, $\nabla$ is the gradient operator with respect to $\vec{x}$, $\hat{T}\big(\vec{X}(t,D),t\big)=D\pa_t+\langle \dot{\vec{X}}(t,D),\hat{\vec{p}}\rangle - \dot{S}(t,D)$, $\dot{S}(t,D)=\dac{dS(t,D)}{dt}$, $\Vert \cdot \Vert$ is the $L_2$-norm, $\langle \cdot, \cdot\rangle$ is the scalar product of vectors, and~$k$ is the non-negative integer. In~particular, \eqref{est1} yields $\hat{\vec{p}}=\hat{\Or}(\sqrt{D})$, $\Delta\vec{x}=\hat{\Or}(\sqrt{D})$. Here, $\hat{F}=\hat{\Or}(D^{\mu})$ means that $\dac{\Vert \hat{F}\varphi\Vert}{\Vert \varphi \Vert}=\Or(D^{\mu})$, $\varphi\in{\mathcal P}_D^t$.

For simplicity, we will omit the parameter $D$ in expressions where it does not cause~confusion.

In view of the estimates \eqref{est1}, the~asymptotic expansion of the coefficient in \mbox{Equation \eqref{int1}} in powers of $\Delta\vec{x}$ in a neighborhood of the trajectory $\vec{x}=\vec{X}(t,D)$ allows us to transform Equation \eqref{int1} to the approximate one with the given accuracy. The residual of the approximate equation has the estimate $\Or\big(D^{\frac{1+M}{2}}\big)$ in the class ${\mathcal P}_D^t$, where $M\geq 2$ is the highest power of $\Delta x_i$ accounted in the expansion. The leading term of asymptotics of the solution to \eqref{int1} is determined by the expansion of up to $\Or(D^{3/2})$. Following~\cite{shapkul21}, the~respective expansions in matrix notations can be written as
\begin{equation}
\begin{gathered}
a(\vec{x},t)=a(\vec{X},t)+a_x \Delta\vec{x} + \dac{1}{2} \Delta\vec{x}^{\top} a_{xx}\Delta \vec{x} + \hat{\Or}(D^{3/2}), \\
b(\vec{x},\vec{y},\vec{z},t)=b(\vec{X},\vec{X},\vec{X},t)+b_x \Delta \vec{x} +b_y \Delta \vec{y} +b_z \Delta \vec{z}  \\
+\dac{1}{2} \Delta\vec{x}^{\top}b_{xx}\Delta\vec{x}+\dac{1}{2}\Delta\vec{y}^{\top}b_{yy}\Delta\vec{y} +\dac{1}{2}\Delta\vec{z}^{\top}b_{zz}\Delta\vec{z}\\
+\Delta\vec{x}^{\top} b_{xy}\Delta\vec{y}  + \Delta\vec{x}^{\top}b_{xz}\Delta\vec{z} +\Delta\vec{y}^{\top}b_{yz}\Delta\vec{z} +\hat{\Or}(D^{3/2}). \end{gathered}
\label{expan1}
\end{equation}
Here, $\vec{X}=\vec{X}(t)=\vec{X}(t,D)$, $\Delta \vec{x}=\vec{x}-\vec{X}(t)$, $\Delta \vec{y}=\vec{y}-\vec{X}(t)$, and~$\vec{z}=\vec{z}-\vec{X}(t)$ are column vectors; $(\cdot)^\top$ is a transposed matrix; $a_x$, $b_x$, $b_y$, and~$b_z$ are row vectors of the form \mbox{$\Big(a_x=\dac{\pa a}{\pa x_i}\Big|_{\vec{x}=\vec{X}(t)}  \Big)$}, $\Big(b_x=\dac{\pa b}{\pa x_i}\Big|_{\vec{x}=\vec{y}=\vec{z}=\vec{X}(t)}\Big)$, row vectors $b_y$, $b_z$ have the analogous form; $a_{xx}$ and $b_{xx}$ are symmetric matrices of the form $a_{xx}=\Big(\dac{\pa^2 a}{\pa x_i \pa x_j}\Big|_{\vec{x}=\vec{X}(t)}\Big)$, \mbox{$b_{xx}=\Big(\dac{\pa^2 b}{\pa x_i \pa x_j}\Big|_{\vec{x}=\vec{y}=\vec{z}=\vec{X}(t)}\Big)$}, matrices $b_{yy}$, $b_{zz}$, $b_{xy}$, $b_{yx}$, $b_{zy}$,$b_{yz}$, and~$b_{xz}$ have the analogous~form.

The key point of the considered approach~\cite{shapkul21} is that the nonlocal nonlinearity enters into the approximate kinetic equation obtained with the help of an asymptotic expansion of \eqref{int1} in the form of the moments of the solution $u(\vec{x},t,D)$, and~dynamical equations that determine the evolution of these moments can be solved separately.

In order to construct the leading term of asymptotics, we need corresponding moments of up to the second order that are defined as follows
\begin{equation}
\begin{gathered}
\sigma_u(t,D)=\dil_{\BR^n}u(\vec{x},t,D)d\vec{x}, \qquad \vec{x}_u(t,D)=\dac{1}{\sigma_u(t,D)}\dil_{\BR^n}\vec{x} u(\vec{x},t,D)d\vec{x}, \\
\big(\alpha_{u,ij}^{(2)}(t,D)\big)=\dac{1}{\sigma_u(t,D)}\dil_{\BR^n}\Delta\vec{x}_i \Delta\vec{x}_j u(\vec{x},t,D)d\vec{x}.
\end{gathered}
\label{mom1}
\end{equation}
Here, $\alpha_{u}^{(2)}(t,D)=\big(\alpha_{u,ij}^{(2)}(t,D)\big)$ is the symmetric matrix of the central second-order moments of the function $u(\vec{x},t,D)$. The~first-order moment will determine the functional parameter $\vec{X}(t,D)$ of the class ${\mathcal P}_D^t$ \eqref{fkpp4} as
\begin{equation}
\vec{x}_u(t,D)=\vec{X}(t,D).
\label{mom2}
\end{equation}

Dynamical equations for the moments \eqref{mom1} are obtained by differentiation with respect to $t$, from~the definitions {\eqref{mom1}}, and~using $\pa_t u(\vec{x},t)$ from \eqref{int1}. Taking into consideration \mbox{expansions \eqref{expan1}} and estimates \eqref{est1} with accuracy of $\Or(D^{3/2})$, we arrive at the following moment equations:
\begin{equation}
\begin{gathered}
\dot{\sigma}_u=\sigma_u\Big(a(\vec{x}_u,t)+\dac{1}{2}\Sp\big[a_{xx}\alpha_u^{(2)}\big]\Big)\\
-\varkappa \sigma_u^3 \Big(b(\vec{x}_u,\vec{x}_u,\vec{x}_u,t)+\dac{1}{2} \Sp\big[(b_{xx}+b_{yy}+b_{zz})\alpha_u^{(2)}\big]\Big), \\
\dot{\vec{x}}_u=(a_x-\varkappa \sigma_u^2 b_x)\alpha_u^{(2)}, \qquad \dot{\alpha}_u^{(2)}=2D\tilde{D}_a(t)\BI_n,
\end{gathered}
\label{eesyst1}
\end{equation}
where $I_n$ is the identity matrix of size $n$, and~the expression $a_{xx} \alpha_u^{(2)}$ implies a product of matrices $a_{xx}$ and $a_u^{(2)}$.

Let us consider Equation \eqref{eesyst1} as the system where the aggregate of moments
\begin{equation}
g_u(t)=\big(\sigma_u(t),\vec{x}_u(t),\alpha_u^{(2)}(t)\big)
\label{gensol1}
\end{equation}
is substituted for the set of independent variables $\big(\sigma(t),\vec{x}(t),\alpha^{(2)}(t)\big)$, and~$\alpha^{(2)}=\big(\alpha_{ij}^{(2)}(t)\big)$ that are not related to the function $u(\vec{x},t)$ in the general case. Thus, the~resulting system can be treated as an independent dynamical system. According to~\cite{shapkul21}, this system is termed the Einstein--Ehrenfest (EE) system of the second order for the nonlocal kinetic Equation \eqref{int1} in the class of TCF \eqref{fkpp4}. The~second order of the system implies that we preserve the terms of order not higher than $\Or(\sqrt{D}^2)$.

Let the general solution of this system be
\begin{equation}
g(t,{\bf C})=\big(\sigma(t,{\bf C}), \vec{x}(t,{\bf C}), \alpha^{(2)}(t,{\bf C})\big),
\label{gensol2}
\end{equation}
where ${\bf C}$ is a set of arbitrary integration constants. Then, the~substitution of the expansion~\eqref{expan1} into Equation \eqref{int1} with the replacement of moments \eqref{gensol1} by the general solution \eqref{gensol2} yields the following linear equation:
\begin{equation}
\hat{\CL}(\vec{x},t,{\bf C})v(\vec{x},t)=0,
\label{ale1}
\end{equation}
where the linear operator $\hat{\CL}(\vec{x},t,{\bf C})$ is given by
\begin{equation}
\begin{gathered}
\hat{\CL}(\vec{x},t,{\bf C})=-\pa_t + D\tilde{D}_a(t) \Delta_x + L(t,{\bf C}) + L_x(t,{\bf C})\Delta \vec{x}+\dac{1}{2}\Delta\vec{x}^{\top} L_{xx}(t,{\bf C}) \Delta \vec{x},\\
L(t,{\bf C})=a\big(\vec{x}(t,{\bf C}),t\big)-\varkappa \sigma^2(t,{\bf C}) \bigg( b\big(\vec{x}(t,{\bf C}),\vec{x}(t,{\bf C}),\vec{x}(t,{\bf C}),t\big)\\
+\dac{1}{2}\Sp\Big[\Big(b_{yy}\big(\vec{x}(t,{\bf C}),\vec{x}(t,{\bf C}),\vec{x}(t,{\bf C}),t\big)+b_{zz}\big(\vec{x}(t,{\bf C}),\vec{x}(t,{\bf C}),\vec{x}(t,{\bf C}),t\big)\Big)\alpha^{(2)}(t,{\bf C})\Big]\bigg),\\
L_x(t,{\bf C})=a_x\big(\vec{x}(t,{\bf C}),t\big)-\varkappa \sigma^2(t,{\bf C}) b_x\big(\vec{x}(t,{\bf C}),\vec{x}(t,{\bf C}),\vec{x}(t,{\bf C}),t\big), \\
L_{xx}(t,{\bf C})=a_{xx}\big(\vec{x}(t,{\bf C}),t\big)-\varkappa \sigma^2(t,{\bf C}) b_{xx}\big(\vec{x}(t,{\bf C}),\vec{x}(t,{\bf C}),\vec{x}(t,{\bf C}),t\big).
\end{gathered}
\label{ale2}
\end{equation}
Equations \eqref{ale1} and \eqref{ale2} in~\cite{shapkul21} are termed the associated linear equation for Equation \eqref{int1}.

Let us pose the Cauchy problem for Equation \eqref{int1} in the class of TCF \eqref{fkpp4}:
\begin{equation}
u(\vec{x},t,D)\big|_{t=0}=\varphi(\vec{x},D)\in\CP^{D}_0, \qquad \CP^{D}_0=\CP^{D}_t\big|_{t=0}.
\label{incond1}
\end{equation}
Next, we impose a restriction on the integration constants ${\bf C}$ involved in the general solutions \eqref{gensol2}. The~restriction is given by the following algebraic condition:
\begin{equation}
g(0,\BFC)=g_\varphi,
\label{alcond1}
\end{equation}
which yields $\BFC=\BFC_\varphi$. Here, $g_\varphi$ is the aggregate of the moments \eqref{gensol1} that is determined by the initial condition $\varphi(\vec{x},D)$ as
\begin{equation}
\begin{gathered}
\sigma_\varphi=\dil_{\BR^n}\varphi(\vec{x},D)d\vec{x}, \qquad \vec{x}_\varphi=\dac{1}{\sigma_\varphi}\dil_{\BR^n} \vec{x} \varphi(\vec{x},D)d\vec{x},\\
\big(\alpha_{\varphi,ij}^{(2)}\big)=\dac{1}{\sigma_\varphi}\dil_{\BR^n} (x_i-x_{\varphi ,i})(x_j - x_{\varphi ,j}\varphi(\vec{x},D)d\vec{x}.
\end{gathered}
\label{incond2}
\end{equation}
Let us consider the Cauchy problem for the associated linear Equations \eqref{ale1} and \eqref{ale2} with the initial condition
\begin{equation}
v(\vec{x},t)\big|_{t=0}=\varphi(\vec{x},D).
\label{incond3}
\end{equation}
According to~\cite{fkppshap18,shapkul21}, the~solution of the Cauchy problem for Equation \eqref{int1} with the initial condition \eqref{incond1} and the solutions of the Cauchy problems \eqref{ale1}, \eqref{incond3} in the class of TCF are related as follows:
\begin{equation}
u(\vec{x},t,D)=v(\vec{x},t,\BFC_\varphi)+\Or(D^{3/2}),
\label{teo1}
\end{equation}
where $\BFC_\varphi$ is given by the algebraic condition \eqref{alcond1}.

\section{Semiclassical asymptotics in a two-dimensional plane-parallel case}
\label{sec3}
In this section, based on the method proposed in the previous section, we will obtain an explicit expression for a family of asymptotic solutions for the special case of \mbox{Equation \eqref{int1}} analogous to the one considered in~\cite{shapkul21}.

Let us consider the problem in the plane orthogonal to the GDT axis. Let {$\vec{x}=(x_1,x_2)\in\BR^2$} be Cartesian coordinates in this plane and the coefficients in \eqref{int1} be given by
\begin{equation}
\begin{gathered}
a(\vec{x},t)=a(t), \qquad b(\vec{x},\vec{y},\vec{z},t)=b(t)p(\vec{x}-\vec{y},\vec{x}-\vec{z},\mu),\\
p(\vec{r}_1,\vec{r}_2,\mu)=\exp\Big[-\dac{\vec{r}_1^2+\vec{r}_2^2}{2\mu^2}\Big],
\end{gathered}
\label{coef1}
\end{equation}
where the functions $a(t)$ and $b(t)$ are assumed to be monotone decreasing and increasing, respectively, and~the parameter $\mu$ characterizes the nonlocality of the nonlinearity kernel $p(\vec{r}_1,\vec{r}_2,\mu)$, $\vec{x}^2=\langle\vec{x},\vec{x}\rangle$.

For functions \eqref{coef1}, Equations \eqref{mom2} and \eqref{eesyst1} yield $\dot{\vec{X}}(t)=0$. The~identity $\dot{\vec{X}}=0$ also holds in a more general case for the problem with the symmetric configuration of a GDT. We choose the origin of coordinates so that $\vec{X}(t)\big|_{t=0}=0$. It leads to $\vec{X}(t)=0$. Ions are usually localized on the GDT axis, which is taken as the origin of coordinates in our case. Then, $\Delta \vec{x}=\vec{x}$.

In the case under consideration, Equation \eqref{eesyst1} reads
\begin{equation}
\dot{\sigma}(t)=a(t)\sigma(t)-\varkappa \beta(t)\sigma^3(t), \qquad \vec{x}(t)=\vec{X}(t)=0, \qquad \alpha^{(2)}(t)=2Dd(t).
\label{eesyst2}
\end{equation}
Here, we denoted
\begin{equation}
\beta(t)=b(t)\Big(1-\dac{4}{\mu^2}D\cdot \Sp [d(t)]\Big), \qquad d(t)=\overline{d}+\BI_2 \dil_0^t \tilde{D}_a(\tau)d\tau, \qquad d(0)=\overline{d},
\label{oboz1}
\end{equation}
where $\overline{d}$ is a diagonal $2\times 2$-matrix, diagonal elements of the matrix $2D\overline{d}$ characterize the degree of localization (dispersion) of the initial axial ion distribution with respect to $x_1$, and~$x_2$, $\BI_2$ is the identity $2\times 2$-matrix.

The general solution of Equation \eqref{eesyst2} is given by
\begin{equation}
\sigma(t)=z^{-2}(t), \qquad z(t)=c^{-1}\varpi(t,0)+2\varkappa \dil_0^t \beta(\tau)\varpi(t,\tau)d\tau,
\label{gensol3}
\end{equation}
where $\varpi(t,\tau)=\exp\Big(-2\dil_{\tau}^t a(\zeta)d\zeta\Big)$ and $c$ is an arbitrary integration constant, $\sigma(0)=c^2$. Note that $\dac{\pa \varpi(t,\tau)}{\pa t}=-2a(t)\varpi(t,\tau)$ and $\varpi(t,t)=1$.

Thus, the~relation \eqref{eesyst2} for $\alpha^{(2)}(t)=\alpha^{(2)}(t,\BFC)$ with $\vec{x}(t)=0$ and the relation \eqref{gensol3} for $\sigma(t)=\sigma(t,\BFC)$ yield the general solution of the EE system with arbitrary integration constant $\BFC$:
\begin{equation}
g(t,\BFC)=\big(\sigma(t,\BFC), \vec{x}(t,\BFC)=0, \alpha^{(2)}(t,\BFC)\big),
\label{gensol4}
\end{equation}
where the set of integration constants reads
\begin{equation}
\BFC=(c^2,0,2D\overline{d}).
\label{intcon1}
\end{equation}

Let us proceed to the construction of the family of particular solutions for the kinetic equation. For~the two-dimensional case under consideration, $\vec{x}\in\BR^2$, with~the coefficients~\eqref{coef1}, Equation \eqref{int1} can be written as
\begin{equation}
\begin{gathered}
\pa_t u(\vec{x},t)=D\tilde{D}_a(t)\Delta_x u(\vec{x},t) + a(t) u(\vec{x},t) \\
-\varkappa b(t) u(\vec{x},t)\dil_{\BR^2}d\vec{y}\dil_{\BR^2}d\vec{z}p(\vec{x}-\vec{y},\vec{x}-\vec{z}) u(\vec{y},t)u(\vec{z},t),
\end{gathered}
\label{kineq1}
\end{equation}
where $\Delta_x$ is a two-dimensional Laplace operator. In~view of \eqref{est1} and \eqref{expan1}, Equation \eqref{kineq1} reads
\begin{equation}
\begin{gathered}
\pa_t u(\vec{x},t)=D\tilde{D}_a(t) \Delta_x u(\vec{x},t) + a(t) u(\vec{x},t)- \\
-\dac{\varkappa}{\mu^2}\sigma_u^2(t) \Big[\mu^2-\Sp \alpha_u^{(2)}(t) - \vec{x}^2\Big]+\Or(D^{3/2}).
\end{gathered}
\label{redeq1}
\end{equation}

Next, we obtain the associated linear Equation \eqref{ale1} from Equation \eqref{redeq1} by the replacement of moments $\sigma_u(t)$, $\alpha_u^{(2)}(t)$ of the required solution $u(\vec{x},t)$ by the general solutions of the dynamical EE system \eqref{gensol4}.

In view of \eqref{eesyst2}, the~relation \eqref{ale2} reads
\begin{equation}
\begin{gathered}
L(t,\BFC)=a(t)-\dac{\varkappa}{\mu^2}\sigma^2(t)\Big(\mu^2-2D\cdot\Sp[d(t)]\Big),\\
L_x(t,\BFC)=0,\\
L_{xx}(t,\BFC)=2\dac{\varkappa}{\mu^2}\sigma^2(t,\BFC)b(t)\BI_2,
\end{gathered}
\label{ale3}
\end{equation}
which allows us to write the associated linear Equation \eqref{ale1} as
\begin{equation}
\begin{gathered}
\hat{\CL}(\vec{x},t,\BFC)v(\vec{x},t)=\bigg(-\pa_t+D\tilde{D}_a(t)\Delta_x+a(t)\\
-\dac{\varkappa}{\mu^2}b(t)\sigma^2(t,\BFC) \Big(\mu^2-2D\cdot\Sp[d(t)]-\vec{x}^2\Big)\bigg)v(\vec{x},t)=0.
\end{gathered}
\label{ale4}
\end{equation}

We are looking for the particular solutions of Equation \eqref{ale4} in the form of the Gaussian~function:
\begin{equation}
v_0(\vec{x},t,\BFC)=N_0 \dac{1}{D}\exp\bigg\{\dac{1}{D}\Big(S(t,D)+\dac{1}{2}\vec{x}^{\top}Q(t)\vec{x}\Big)+\phi(t,D)\bigg\},
\label{vac1}
\end{equation}
where $N_0$ is a normalization constant that is related to the initial number of ions. The~multiplier $\dac{1}{D}$ is introduced for convenience. Note that this multiplier does not contradict the ansatz \eqref{fkpp4} since it can be included in the function $S(t,D)$ as the summand $(-D\ln D)$. The~term $(-D\ln D)\to 0$ as $D\to 0$, so it does not violate the regularity of $S(t,D)$ with respect to $\sqrt{D}$ as $D\to 0$.

The substitution of \eqref{vac1} into \eqref{ale4} yields
\begin{equation}
\begin{gathered}
\dot{S}=D\bigg[a(t)-\varkappa b(t) \sigma^2(t) \Big(1-\dac{2D}{\mu^2}\Sp[d(t)]\Big)\bigg], \qquad \phi(t,D)=\dil_0^t  \tilde{D}_a(\tau) \Sp Q(\tau)d\tau,\\
\dac{1}{2}\dot{Q}=\tilde{D}_a(t)Q^2+D\dac{\varkappa}{\mu^2}b(t)\sigma^2(t)\BI_2.
\end{gathered}
\label{sfunc1}
\end{equation}
Equation \eqref{sfunc1} determines the functions $S(t,D)$ and $\phi(t,D)$ through quadratures:
\begin{equation}
\begin{gathered}
S(t,D)=D\dil_0^t d\tau \bigg[ a(\tau)-\varkappa b(\tau) \sigma^2(\tau) \Big(1-\dac{2D}{\mu^2}\Sp[d(\tau)]\Big) \bigg],\\
\phi(t,D)=\dil_0^t d\tau \tilde{D}_a \Sp Q(\tau).
\end{gathered}
\label{sfunc2}
\end{equation}

The matrix Riccati equation in \eqref{sfunc1} can be represented as the linear system by the~substitution:
\begin{equation}
Q(t)=B(t)C^{-1}(t),
\label{ric1}
\end{equation}
where $B(t)$ and $C(t)$ are nondegenerate matrices that satisfy the following matrix linear system of differential equations:
\begin{equation}
\left\{\begin{array}{l}
\dot{C}(t)=-2\tilde{D}_a(t)B(t),\cr
\dot{B}(t)=2\dac{\varkappa}{\mu^2}D b(t)\sigma^2(t)C(t).
\end{array}
\right.
\label{var1}
\end{equation}

\section{Countable set of solutions to the associated linear equation}
\label{sec4}
In this section, we describe the approach to constructing the family of solutions of the associated linear Equation \eqref{ale4} based on solutions of the system \eqref{var1}.

Denote $\tilde{b}(t)=\dac{\varkappa}{\mu^2}D b(t) \sigma^2(t)$. Since the coefficients in the system \eqref{var1} are scalar, the~matrices $B(t)$ and $C(t)$ can be sought in the diagonal form:
\begin{equation}
B(t)=\diag\big\{W_1(t), W_2(t)\big\}, \qquad C(t)=\diag\big\{Z_1(t),Z_2(t)\big\},
\label{ric2}
\end{equation}
where $W_i(t)$ and $Z_i(t)$, $i,j=1,2$, are scalar~functions.

Note that the matrix $Q(t)$ is also diagonal is this case:
\begin{equation}
Q(t)=\diag\Big\{\dac{W_1(t)}{Z_1(t)},\dac{W_2(t)}{Z_2(t)}\Big\}
\label{ric3}
\end{equation}

For the functions in \eqref{ric2}, the~system \eqref{var1} reads
\begin{equation}
\dot{Z}_i(t)=-2\tilde{D}_a(t)W_i(t), \qquad \dot{W}_i(t)=2\tilde{b}(t)Z_i(t).
\label{var2}
\end{equation}

The equations for functions $\big(W_1(t),Z_1(t)\big)$ and $\big(W_2(t),Z_2(t)\big)$ are identical. Hence, these functions can differ only due to their different initial condition. Let us consider the system of equations for the functions $W(t)$ and $Z(t)$,
\begin{equation}
\dot{Z}(t)=-2\tilde{D}_a(t)W(t), \qquad \dot{W}(t)=2\tilde{b}(t)Z(t),
\label{var3}
\end{equation}
whose solutions for different initial conditions determine the matrices $B(t)$ and $C(t)$ and, correspondingly, the~matrix $Q(t)$. The~system \eqref{var3} was coined ``the variational system''\, in~\cite{fkppshap18}.

The system \eqref{var3} has two linearly independent solutions. Let us denote them by the following formulae:
\begin{equation}
\Big(W^{(\pm)}(t),Z^{(\pm)}(t)\Big).
\label{varsol1}
\end{equation}
These solutions are determined by the following initial conditions:
\begin{equation}
W^{(+)}(0)=\beta>0, \qquad W^{(-)}(0)=-\beta, \qquad Z^{(\pm)}(0)=1.
\label{varsol2}
\end{equation}
For the functions $\Big(W_i^{\pm}(t),Z_i^{\pm}(t)\Big)$, $i,j=1,2$, the~numbers $\beta_i$ that determine the initial conditions by
\begin{equation}
W^{(+)}_i(0)=\beta_i>0, \qquad W^{(-)}_i(0)=-\beta_i, \qquad Z_i^{(\pm)}(0)=1
\label{varsol3}
\end{equation}
can be different in a general~case.

Now, we can construct the family of solutions to Equation \eqref{ale4} based on the particular solution \eqref{vac1}. For~this purpose, we use the well-known quantum mechanics method widely used in various problems~\cite{Manko79,perelomov86,obukhov22}. 

Denote the two-dimensional symplectic identity matrix as $J=\begin{pmatrix}0 & -1 \cr 1 & 0\end{pmatrix}$. Introduce the two-dimensional column vectors
\begin{equation}
a^{(\pm)}(t)=\Big(W^{(\pm)}(t),Z^{(\pm)}(t)\Big)^{\top}
\label{sym1}
\end{equation}
and the skew scalar product
\begin{equation}
\big\{a^{(-)}(t),a^{(+)}(t)\big\}=\big(a^{(-)}(t)\big)^{\top} J a^{(+)}(t)=Z^{(-)}(t)W^{(+)}(t)-Z^{(+)}(t)W^{(-)}(t),
\label{sym2}
\end{equation}
which is a conserved quantity in view of \eqref{var3}:
\begin{equation}
\begin{gathered}
\dac{d}{dt}\big\{a^{(-)}(t),a^{(+)}(t)\big\}=\dot{Z}^{(-)}(t)W^{(+)}(t)+Z^{(-)}\dot{W}^{(+)}(t)-\dot{Z}^{(+)}(t)W^{(-)}(t)\\
-Z^{(+)}(t)\dot{W}^{(-)}(t)=-2\tilde{D}_a(t)W^{(-)}(t)W^{(+)}(t)+Z^{(-)}(t)2\tilde{b}(t)Z^{(+)}(t)\\
+2\tilde{D}_a(t)W^{(+)}(t)W^{(-)}(t)-Z^{(+)}(t)2\tilde{b}(t)Z^{(+)}(t)=0.
\end{gathered}
\label{chern1}
\end{equation}
Therefore, for~$t=0$, we have
\begin{equation}
\big\{a^{(-)}(0),a^{(+)}(0)\big\}=Z^{(-)}(0)W^{(+)}(0)-Z^{(+)}(0)W^{(-)}(0)=2\beta,
\label{chern2}
\end{equation}
i.e.,
\begin{equation}
\big\{a^{(-)}(t),a^{(+)}(t)\big\}=2\beta.
\label{sym3}
\end{equation}

Following the general Maslov complex germ method, the~linear $two$-dimensional space with the basis vectors $a_1^{(-)}(t)$ and $a_2^{(-)}(t)$ is called the germ $r^{2}_t$. In~our case, the~germ is chosen to be real, since we seek the real solutions to \eqref{kineq1}. The~pair $\big(\Lambda^{0}_t,r^{2}_t\big)$, where $\Lambda^{0}_t$ is the trajectory (the time-dependent $0$-dimensional manifold) $\vec{x}=\vec{X}(t)$, determines the set of asymptotic solutions to Equation \eqref{kineq1}. In~order to construct such solutions, we present the symmetry operators associated with the vectors $a_1^{(-)}(t)$, $a_2^{(-)}(t)$ and with the vectors $a_1^{(+)}(t)$, $a_2^{(+)}(t)$.

Define the operators
\begin{equation}
\begin{gathered}
\hat{a}^{(-)}(t)=-N_a\Big(Z^{(-)}(t)D\pa_x-W^{(-)}(t)x\Big),\\
\hat{a}^{(+)}(t)=N_a \Big( Z^{(+)}(t)D\pa_x-W^{(+)}(t)x \Big).
\end{gathered}
\label{sym4}
\end{equation}
These operators satisfy the following commutation relations:
\begin{equation}
\begin{gathered}
\big[\hat{a}^{(-)}(t),\hat{a}^{(-)}(t)\big]=\big[\hat{a}^{(+)}(t),\hat{a}^{(+)}(t)\big]=0,\\
\big[\hat{a}^{(-)}(t),\hat{a}^{(+)}(t)\big]=N_a^2 \Big[-Z^{(-)}(t)D\pa_x+W^{(-)}(t)x, Z^{(+)}(t)D\pa_x-W^{(+)}(t)x\Big]\\
=N_a^2D\Big(Z^{(-)}(t)W^{(+)}(t)-Z^{(+)}(t)W^{(-)}(t)\Big)=N_a^2D\cdot2\beta.
\end{gathered}
\label{sym5}
\end{equation}
We define the normalization multiplier as $N_a=\dac{1}{\sqrt{2\beta D}}$. Then, we have
\begin{equation}
\begin{gathered}
\hat{a}^{(-)}(t)=-\dac{1}{\sqrt{2\beta D}}\Big(Z^{(-)}(t)D\pa_x-W^{(-)}(t)x\Big),\\
\hat{a}^{(+)}(t)=-\dac{1}{\sqrt{2\beta D}}\Big(Z^{(+)}(t)D\pa_x-W^{(+)}(t)x\Big),
\end{gathered}
\label{sym6}
\end{equation}
and
\begin{equation}
\begin{gathered}
\big[\hat{a}^{(-)}(t),\hat{a}^{(+)}(t)\big]=1.
\end{gathered}
\label{sym7}
\end{equation}

For the diagonal matrix $Q(t)$ \eqref{ric3}, the~solution \eqref{vac1} can be written as
\begin{equation}
v_0(\vec{x},t,\BFC)=N_0\dac{1}{D}\exp\bigg\{\dac{1}{D}\Big[S(t,D)+\dac{1}{2}\big(Q_1(t)x_1^2+Q_2(t)x_2^2\big)\Big]+\phi(t,D)\bigg\}.
\label{vac2}
\end{equation}
Here, $Q_i(t)=\dac{W_i^{(-)}(t)}{Z_i^{(-)}(t)}$, and~the functions $\big(W_i^{(\pm)},Z_i^{(\pm)}\big)$ are solutions of the system \eqref{var3} with initial conditions \eqref{varsol3}.

For the two-dimensional case, the~operators \eqref{sym6} read
\begin{equation}
\begin{gathered}
\hat{a}_i^{(-)}(t)=-\dac{1}{\sqrt{2\beta_i D}}\Big(Z_i^{(-)}(t)D\pa_{x_i}-W_i^{(-)}(t)x_i\Big),\\
\hat{a}_i^{(+)}(t)=-\dac{1}{\sqrt{2\beta_i D}}\Big(Z_i^{(+)}(t)D\pa_{x_i}-W_i^{(+)}(t)x_i\Big),
\end{gathered}
\label{sym8}
\end{equation}
and the commutators are as follows:
\begin{equation}
\big[\hat{a}_i^{(-)}(t),\hat{a}_j^{(+)}(t)\big]=\delta_{ij}.
\label{sym9}
\end{equation}

It can be shown that the operators $\hat{a}_i^{(-)}(t)$ nullify the function $v_0(\vec{x},t,\BFC)$ of the form~\eqref{vac2}
\begin{equation}
\hat{a}_i^{(-)}(t)v_0(\vec{x},t,\BFC)=0,
\label{sym10}
\end{equation}
and that operators $\hat{a}^{(\pm)}(t)$ commute with the operator $\hat{\CL}(\vec{x},t,\BFC)$ of Equation \eqref{ale4}:
\begin{equation}
\big[\hat{\CL}(\vec{x},t,\BFC),\hat{a}_i^{(\pm)}(t)\big]=0.
\label{sym11}
\end{equation}

It means that the operators $\hat{a}_i^{(\pm)}(t)$ are the symmetry operators for Equation \eqref{ale4}. Hence, the~action of the operators $\hat{a}_i^{(+)}(t)$ on $v_0(\vec{x},t,\BFC)$ generates the family of new solutions to Equation \eqref{ale4}. Let us define this family of solutions by
\begin{equation}
v_n(\vec{x},t,\BFC)=\dac{N_n}{N_0}\dac{1}{n!}\big(\hat{a}_1^{(+)}(t)\big)^{n_1}\big(\hat{a}_2^{(+)}(t)\big)^{n_2}v_0(\vec{x},t,\BFC).
\label{exc1}
\end{equation}
Here, $n=(n_1,n_2)$ is the two-dimensional multi-index, $n_i\in\{0,1,2,\ldots\}$, $|n|=n_1+n_2$, $n!=n_1!n_2!$, $N_n$ are normalization~coefficients.

The solutions to \eqref{exc1} can be written in the explicit form. Define
\begin{equation}
\xi_i=\sqrt{\dac{\beta_i}{D Z_i^{(-)}(t)Z_i^{(+)}(t)}}x_i.
\label{ksi1}
\end{equation}
In view of the formula for the Hermitian polynomials~\cite{hermite}
\begin{equation}
H_{n_i}(\xi_i)=(-1)^{n_i}\exp(\xi_i^2)(\pa_{\xi_i})^{n_i}\exp(-\xi_i^2),
\label{herm1}
\end{equation}
the relation \eqref{exc1} reads
\begin{equation}
\begin{gathered}
v_n(\vec{x},t,\BFC)=\dac{N_n}{N_0}\dac{(-1)^{|n|}}{\sqrt{2^{|n|}n!}}\bigg(\dac{Z_1^{(+)}(t)}{Z_1^{(-)}(t)}\bigg)^{\frac{n_1}{2}}
\bigg(\dac{Z_2^{(+)}(t)}{Z_2^{(-)}(t)}\bigg)^{\frac{n_2}{2}}H_{n_1}(\xi_1)H_{n_2}(\xi_2)v_0(\vec{x},t,\BFC).
\end{gathered}
\label{exc2}
\end{equation}

The set \eqref{exc2} is the parametric family of solutions to the associated linear Equation \eqref{ale4}. Our next task is to find a countable set of asymptotic solutions to the kinetic Equation \eqref{kineq1} among these solutions with the free parameter $\BFC$.

\section{Algebraic conditions and solutions of the kinetic equation}
\label{sec5}
In this section, we construct solutions $u_n(\vec{x},t)$ to Equation \eqref{redeq1} with help of the solutions $v_n(\vec{x},t,\BFC)$ \eqref{exc2} to the associated linear equation \eqref{ale4}. For~this, we impose the algebraic conditions \eqref{alcond1} and \eqref{incond2} on arbitrary integration constants $\BFC=(c^2,0,2D\overline{d})$ that are included in the solution of the EE system. Then, we have
\begin{equation}
u_n(\vec{x},t)=v_n(\vec{x},t,\BFC_n),
\label{cons1}
\end{equation}
where
\begin{equation}
\BFC_n=(c_n^2,0,2D\overline{d}_n), \qquad n=(n_1,n_2).
\label{cons2}
\end{equation}
The functions $u_n(\vec{x},t)$ are the leading terms of asymptotics for solutions to the original nonlocal kinetic Equation \eqref{kineq1} with accuracy of $\Or(D^{3/2})$.

Next, we obtain the constants \eqref{cons2} in an explicit form. Let us write the solutions to  \eqref{exc2} in the form
\begin{equation}
\begin{gathered}
v_n(\vec{x},t,\BFC)=\Upsilon_n(t) H_{n_1}(\xi_1)H_{n_2}(\xi_2)\dac{1}{D}\exp\Big\{\dac{1}{2D}\big(Q_1(t)x_1^2+Q_2(t)x_2^2\big)\Big\},\\
\Upsilon_n(t)=\dac{(-1)^{|n|}}{\sqrt{2^{|n|}n!}}\bigg(\dac{Z_1^{(+)}(t)}{Z_1^{(-)}(t)}\bigg)^{\frac{n_1}{2}}
\bigg(\dac{Z_2^{(+)}(t)}{Z_2^{(-)}(t)}\bigg)^{\frac{n_2}{2}}N_n
\exp\Big\{\dac{1}{D}S(t,D)+\phi(t,D)\Big\},\\
Q_i(t)=\dac{W_i^{(-)}}{Z_i^{(-)}(t)}, \qquad \xi_i=\sqrt{\dac{\beta_i}{D Z_i^{(-)}(t)Z_i^{(+)}(t)}}x_i.
\end{gathered}
\label{genfun1}
\end{equation}
In order to obtain the moments for the whole set of solutions $v_n(\vec{x},t,\BFC)$, we use the representation of the Hermitian polynomial through the generating function~\cite{hermite}:
\begin{equation}
\exp[2xt-t^2]=\sum_{k=0}^{\infty}H_k(x)\dac{t^k}{k!}.
\label{genfun2}
\end{equation}
Then, the~generating function $V(\vec{x},t,t_1,t_2,\BFC)$ for solutions \eqref{genfun1} is given as follows,
\begin{equation}
\begin{gathered}
V(\vec{x},t,t_1,t_2,\BFC)=\exp\bigg\{2\sqrt{\dac{\beta_1}{D Z_1^{(-)}(t)Z_1^{(+)}(t)}}t_1 x_1+2\sqrt{\dac{\beta_2}{D Z_2^{(-)}(t)Z_2^{(+)}(t)}}t_2 x_2\bigg\}\\
\times \dac{1}{D}\exp\bigg\{\dac{1}{2D}\bigg(\dac{W_1^{(-)}(t)}{Z_1^{(-)}(t)}x_1^2+\dac{W_2^{(-)}(t)}{Z_2^{(-)}(t)}x_2^2\bigg)\bigg\},\\
V(\vec{x},t,t_1,t_2,\BFC)=\sum_{n_1=0}^{\infty}\sum_{n_2=0}^{\infty}\Upsilon_n^{-1}(t)v_n(\vec{x},t,\BFC)\dac{t_1^{n_1}t_2^{n_2}}{n!}.
\end{gathered}
\label{genfun3}
\end{equation}
From \eqref{mom1}, one can see that $\sigma_u(t)$ and $\alpha^{(2)}_u(t)\cdot\sigma_u(t)$ are linear functionals with respect to $u(\vec{x},t)$. This property allows us to obtain moments $\sigma_n(t)$ and $\alpha^{(2)}_n(t)\cdot\sigma_n(t)$ of the functions $v_n(\vec{x},t,\BFC)$ \eqref{genfun1} with the help of respective generating functions derived from \eqref{genfun3}. Introduce the following functions:
\begin{equation}
\begin{gathered}
\Omega(t,t_1,t_2)=\dil_{\BR^2}V(\vec{x},t,t_1,t_2,\BFC)d\vec{x},\\
\Omega(t,t_1,t_2)=\sum_{n_1=0}^{\infty}\sum_{n_2=0}^{\infty}\Upsilon^{(-1)}_n(t)\sigma_n(t)\dac{t_1^{n_1}t_2^{n_2}}{n!},\\
A_{ii}(t,t_1,t_2)=\dil_{\BR^2}x_i^2V(\vec{x},t,t_1,t_2,\BFC)d\vec{x},\qquad i=1,2,\\
A_{ii}(t,t_1,t_2)=\sum_{n_1=0}^{\infty}\sum_{n_2=0}^{\infty}\Upsilon_n^{-1}(t)\alpha_{ii,n}^{(2)}(t) \sigma_n(t)\dac{t_1^{n_1}t_2^{n_2}}{n!}.
\end{gathered}
\label{genfun4}
\end{equation}
Here, the~subscripts $ii$ indicate the number of a matrix~element.

Straightforward calculations of integral in the formulae \eqref{genfun4} yield
\vspace{-9pt}
\begin{equation}
\begin{gathered}
\Omega(t,t_1,t_2)=\dac{2\pi\sqrt{Z_1^{(-)}(t)Z_2^{(-)}}}{\sqrt{W_1^{(-)}(t)W_2^{(-)}}}
\exp\bigg[-2\bigg(\dac{\beta_1 t_1^2}{Z_1^{(+)}(t)W_1^{(-)}(t)}+\dac{\beta_2 t_2^2}{Z_2^{(+)}(t)W_2^{(-)}(t)}\bigg)-t_1^2-t_2^2\bigg],\\
A_{11}(t,t_1,t_2)=\dac{2\pi D\sqrt{\big(Z_1^{(-)}(t)\big)^5 Z_2^{(-)}(t)}}{\sqrt{\big(W_1^{(-)}(t)\big)^5 W_2^{(-)}(t)}}\bigg(-\dac{W_1^{(-)}(t)}{Z_1^{(-)}(t)}+4t_1^2\dac{\beta_1}{Z-1^{(-)}(t)Z_1^{(+)}(t)}\bigg)\\
\times\exp\bigg[-2\bigg(\dac{\beta_1 t_1^2}{Z_1^{(+)}(t)W_1^{(-)}(t)}+\dac{\beta_2 t_2^2}{Z_2^{(+)}(t)W_2^{(-)}(t)}\bigg)-t_1^2-t_2^2\bigg].
\end{gathered}
\label{genfun5}
\end{equation}
The formula for $A_{22}(t,t_1,t_2)$ can be obtained from one for $A_{11}(t,t_1,t_2)$ \eqref{genfun5} by the formal interchanging $1\leftrightarrow 2$ in all subscripts. In~view of \eqref{genfun4}, the~expansion of functions \eqref{genfun5} in powers of $t_1$, $t_2$ yields the following expression for moments:
\begin{equation}
\begin{gathered}
\sigma_{(2n_1,2n_2)}(t)=\Upsilon_{(2n_1,2n_2)}(t)\dac{2\pi\sqrt{Z_1^{(-)}(t)Z_2^{(-)}(t)}}{\sqrt{W_1^{(-)}(t)W_2^{(-)}(t)}}\dac{(2n_1)!(2n_2)!}{n_1!n_2!}\\
\times\bigg[-\dac{2\beta_1}{Z_1^{(+)}(t)W_1^{(-)}(t)}-1\bigg]^{n_1}\bigg[-\dac{2\beta_2}{Z_2^{(+)}(t)W_2^{(-)}(t)}-1\bigg]^{n_2},\\\\
\sigma_{(2n_1+1,2n_2)}(t)=\sigma_{(2n_1,2n_2+1)}(t)=\sigma_{(2n_1+1,2n_2+1)}(t)=0,\\\\
\alpha_{11,(2n_1,2n_2)}^{(2)}(t)=-D\dac{Z_1^{(-)}(t)}{W_1^{(-)}(t)}\Bigg\{1+\bigg[\dac{2\beta_1}{Z_1^{(+)}(t)W_1^{(-)}(t)}+1\bigg]^{-1}\dac{4\beta_1}{W_1^{(-)}(t)Z_1^{(+)}(t)}n_1\Bigg\},\\
\alpha_{22,(2n_1,2n_2)}^{(2)}(t)=-D\dac{Z_2^{(-)}(t)}{W_2^{(-)}(t)}\Bigg\{1+\bigg[\dac{2\beta_2}{Z_2^{(+)}(t)W_2^{(-)}(t)}+1\bigg]^{-1}\dac{4\beta_1}{W_2^{(-)}(t)Z_2^{(+)}(t)}n_2\Bigg\}.
\end{gathered}
\label{genfun6}
\end{equation}
Thus, among~the obtained solutions $v_{(n_1,n_2)}(\vec{x},t,\BFC)$ to the associated linear Equation \eqref{ale4}, only those ones with even indices $n_1$, $n_2$ generate the asymptotic solutions to the kinetic Equation \eqref{kineq1}.

Note that functions \eqref{genfun6} are particular solutions to the EE system \eqref{eesyst2}. Hence, the~solutions to the EE system \eqref{eesyst2} are determined by the solutions to the variational system \eqref{var2}. It is a corollary of the fact shown in \protect{\cite{shapkul21}} that the leading term of asymptotics for the function $u(\vec{x},t)$ uniquely defines the functions $\sigma(t)$, $\alpha^{(2)}(t)$ within accuracy of $\Or(D^{3/2})$.

Substituting $t=0$ into the formulae \eqref{genfun6} and taking into account \eqref{varsol3} and \eqref{sfunc2}, we obtain the following initial conditions for moments included in the integration constants $\BFC_n$:
\begin{equation}
\begin{gathered}
\sigma_{(2n_1,2n_2)}(0)=\dac{N_{(2n_1,2n_2)}}{2^{n_1+n_2-1}}\dac{\pi}{\sqrt{\beta_1\beta_2}}\dac{\sqrt{(2n_1)!(2n_2)!}}{n_1!n_2!},\\
\alpha_{11,(2n_1,2n_2)}^{(2)}(0)=\dac{D}{\beta_1}(1+4n_1), \qquad \alpha_{22,(2n_1,2n_2)}^{(2)}(0)=\dac{D}{\beta_2}(1+4n_2),
\end{gathered}
\label{cons3}
\end{equation}
while the integration constants \eqref{cons2} read
\begin{equation}
\begin{gathered}
c^2_{(2n_1,2n_2)}=\sigma_{(2n_1,2n_2)}(0), \qquad \overline{d}_{(2n_1,2n_2)}=\diag\Big(\dac{1}{2\beta_1}(1+4n_1),\dac{1}{2\beta_2}(1+4n_2)\Big).
\end{gathered}
\label{cons4}
\end{equation}

Thus, the~functions \eqref{cons1} with the constants $\BFC_{(2n_1,2n_2)}=\big(c^2_{(2n_1,2n_2)},0,2D\overline{d}_{(2n_1,2n_2)}\big)$ of the form \eqref{cons3} and \eqref{cons4} determine a countable set of solutions to Equation \eqref{redeq1} that are the leading terms of semiclassical asymptotics for the kinetic Equation \eqref{kineq1}.

Note that the solution in Equation \eqref{exc2} for $n\neq 0$ changes its sign in the space $\vec{x}\in\BR^2$ due to the properties of Hermitian polynomials. For~the main physical applications, the~function $u(\vec{x},t)$ is positive definite (e.g., it corresponds to the ion concentration in the model of the MVAM kinetics). Therefore, the~functions $v_n(\vec{x},t,\BFC)$ are rather of interest not by themselves but as a basis for the expansion of positive definite functions. In~such interpretation, the~function $v_n(\vec{x},t,\BFC)$ is a ``mode''\, of~the physical state involving $v_n(\vec{x},t,\BFC)$ in its expansion. In~order to clarify the meaning of this statements, let us describe the nonlinear superposition principle for semiclassical solutions of Equation \eqref{kineq1}.

The functions $v_n(\vec{x},t)$ \eqref{genfun1} can be written as
\begin{equation}
\begin{gathered}
v_n(\vec{x},t)=\Upsilon_n(t)\Psi_n(\vec{x},t), \qquad n=(n_1,n_2),\\
\Psi_n(\vec{x},t)=H_{n_1}(\xi_1)H_{n_2}(\xi_2)\dac{1}{D}
\exp\bigg\{\dac{1}{2D}\bigg(\dac{W_1^{(-)}(t)}{Z_1^{(-)}(t)}x_1^2+\dac{W_2^{(-)}(t)}{Z_2^{(-)}(t)}x_2^2\bigg)\bigg\},\\
\xi_i=\sqrt{\dac{\beta_i}{DZ_i^{(-)}(t)Z_i^{(+)}(t)}}x_i.
\end{gathered}
\label{suppos1}
\end{equation}

The substitution of $t=0$ into \eqref{suppos1} yields
\begin{equation}
\psi_n(\vec{x})=\Psi_n(\vec{x},0)=H_{n_1}\bigg(x_1\sqrt{\dac{\beta_1}{D}}\bigg)H_{n_2}\bigg(x_2\sqrt{\dac{\beta_2}{D}}\bigg)
\dac{1}{D}\exp\Big\{-\dac{1}{2D}(\beta_1 x_1^2+\beta_2 x_2^2)\Big\}.
\label{suppos2}
\end{equation}

It can be seen that the following orthogonality condition holds for the functions $\psi_n(\vec{x})$:
\begin{equation}
\dil_{\BR^2}\psi_n(\vec{x})\psi_m(\vec{x})d\vec{x}=\dac{2^{|n|}n!\pi}{D\sqrt{\beta_1\beta_2}}\delta_{n_1 m_1}\delta_{n_2 m_2}, \qquad m=(m_1,m_2).
\label{orth1}
\end{equation}
Let us expand the initial condition \eqref{incond3} in functions $\psi_n(\vec{x})$:
\begin{equation}
\varphi(\vec{x})=\sum_{n_1=0}^{\infty} \sum_{n_2=0}^{\infty} k_n \psi_n(\vec{x}).
\label{suppos3}
\end{equation}
Then, the~initial condition $\varphi(\vec{x})$ corresponds to the following solution to the associated linear Equation \eqref{ale4}:
\begin{equation}
\begin{gathered}
v(\vec{x},t,\BFC)=\exp\Big\{\dac{1}{D}S(t)+\phi(t)\Big\}\sum_{n_1=0}^{\infty}\sum_{n_2=0}^{\infty}
k_n \bigg(\dac{Z_1^{(+)}(t)}{Z_1^{(-)}(t)}\bigg)^{\frac{n_1}{2}}\bigg(\dac{Z_2^{(+)}(t)}{Z_2^{(-)}(t)}\bigg)^{\frac{n_2}{2}}\Psi_n(\vec{x},t),\\
v(\vec{x},t,\BFC)\Big|_{t=0}=\sum_{n_1=0}^{\infty} \sum_{n_2=0}^{\infty} k_n \psi_n(\vec{x}),\qquad \Psi_n(\vec{x},0)=\psi_n(\vec{x}).
\end{gathered}
\label{suppos4}
\end{equation}

In order to obtain the asymptotic solution of the original kinetic Equation \eqref{kineq1}, we must impose the algebraic condition $\BFC=\BFC_{\varphi}$. The~integration constants $\BFC_{\varphi}$ for the function $\varphi(\vec{x})$ \eqref{suppos3} are given by
\begin{equation}
\begin{gathered}
\sigma(0)=\sum_{n_1=0}^{\infty}\sum_{n_2=0}^{\infty}k_{(2n_1,2n_2)}\dac{2\pi}{\sqrt{\beta_1\beta_2}}\bigg(-\dac{1}{\sqrt{2}}\bigg)^{n_1+n_2}\dac{\sqrt{(2n_1)!(2n_2)!}}{\sqrt{n_1!n_2!}}, \qquad \sigma(0)\neq 0,\\
\alpha_{ii}^{(2)}(0)=\dac{1}{\sigma(0)}\dac{D}{\beta_i}\sum_{n_1=0}^{\infty}\sum_{n_2=0}^{\infty}
k_{(2n_1,2n_2)}\dac{2\pi}{\sqrt{\beta_1\beta_2}}\bigg(-\dac{1}{\sqrt{2}}\bigg)^{n_1+n_2}\dac{\sqrt{(2n_1)!(2n_2)!}}{\sqrt{n_1!n_2!}}(1+4n_i),
\end{gathered}
\label{suppos5}
\end{equation}
where $i=1,2$.

Thus, a~set of the integration constants $\BFC_{\varphi}$ is determined by a set of expansion coefficients of an initial condition $\varphi(\vec{x})$ with respect to functions $\psi_n(\vec{x},t)$, i.e.,~expansion coefficients nonlinearly determine the asymptotic solution $u(\vec{x},t)=v(\vec{x},t,\BFC)+\Or(D^{3/2})$ to the nonlinear kinetic Equation \eqref{kineq1} with the initial condition $u(\vec{x},t)\Big|_{t=0}=\varphi(\vec{x})$. The~nonlinearity of this superposition principle is caused by the nonlinear dependence of the functions included in \eqref{suppos4} (in particular, the~solutions of the variational system) on the integration constants in \eqref{suppos5}.

Although the solutions of the associated linear equation $v_{(2n_1,2n_2+1)}(\vec{x},t)$, $v_{(2n_1+1,2n_2)}(\vec{x},t)$, $v_{(2n_1+1,2n_2+1)}(\vec{x},t)$ do not correspond to any asymptotic solutions of the kinetic equation in themselves, the~functions $\Psi_{(2n_1,2n_2+1)}(\vec{x},t)$, $\Psi_{(2n_1+1,2n_2)}(\vec{x},t)$, and~$\Psi_{(2n_1+1,2n_2+1)}(\vec{x},t)$ given by \eqref{suppos1} enter into the expansion \eqref{suppos4} subject to $\sigma(0)\neq 0$. Thus, we can construct asymmetrical solutions with respect to spatial variables (not odd and not even) to the nonlinear kinetic equation with the help of the nonlinear superposition~principle.

Note that our functions $v_n(\vec{x},t,\BFC)$ contain two free positive parameters, $\beta_1$ and $\beta_2$. The~change of these parameters yields us a new family of the asymptotic solutions \eqref{cons1} or a new basis \eqref{suppos1} for the nonlinear superposition principle. For~example, the~solution $v_0(\vec{x},t,\BFC_0)$ is invariant under the interchanging $x_1\leftrightarrow x_2$ for $\beta_1=\beta_2$ and is not invariant for $\beta_1\neq\beta_2$. The~parameters $\beta_1$, $\beta_2$ determine the localization area of the functions $v_n(\vec{x},t,\BFC)$ and $\Psi_n(\vec{x},t,\BFC)$ with respect to $x_1$, $x_2$. The~nonlinear superposition principle can be applied to the given initial condition under any positive $\beta_1$, $\beta_2$. However, the~more the localization area of the initial condition differs from the localization area of the functions $\Psi_n(\vec{x},t,\BFC)$ determined by $\beta_1$, $\beta_2$, the~slower the series \eqref{suppos3} converges. The~slow convergence of the series \eqref{suppos3} leads to the slow convergence of the series \eqref{suppos4} for $t>0$. Yet, it is not clear how to compare the localization area of two multipeak functions in a general case. It can be completed in the particular case though, which is illustrated in the next section. We can draw an analogy for the parameters $\beta_1$, $\beta_2$ with the scaling factor (dilations) of the wavelets~\cite{daubechies92}, since both the functions {\eqref{suppos1}} and their Fourier transform with respect to $\vec{x}$ are localized functions at each fixed $t$ with the localization area determined by $\beta_1$, $\beta_2$. The proposed rule of thumb for the choice of $\beta_1$, $\beta_2$ corresponds to the value of dilations for the peak of the wavelet~image.

\section{Example of the semiclassical two-dimensional distribution}
\label{sec6}
In this section, we construct the semiclassical solutions to Equation \eqref{kineq1} with the initial~condition
\begin{equation}
\varphi(\vec{x})=\dac{N}{D}\bigg(\exp\bigg[-\dac{\vec{x}^2}{2\gamma_1 D}\bigg]-\varepsilon\cdot \exp\bigg[-\dac{\vec{x}^2}{2\gamma_2 D}\bigg]\bigg),
\label{exam1}
\end{equation}
where $\gamma_1>\gamma_2>0$, $0\leq \varepsilon \leq 1$. For~such constraints for the parameters $\gamma_1$, $\gamma_2$, $\varepsilon$, and~$N>0$, the~function $\varphi(\vec{x})$ does not take on negative values. If~$\varepsilon>\dac{\gamma_2}{\gamma_1}$, then the initial distribution~\eqref{exam1} has a minimum at the center $\vec{x}=0$. The~minimum of the ion distribution described by Equation \eqref{kineq1} is physically realizable by the addition of hydrogen into the metal vapor active medium (by creating the so-called kinetically enhanced active medium)~\cite{marshall04,boichenko03,Boichenko2011189}.

Let us apply the nonlinear superposition principle to the initial condition \eqref{exam1}. Note~that
\begin{equation}
\begin{gathered}
h(\vec{x},t_1,t_2)=\exp\bigg\{2\sqrt{\dac{\beta_1}{D}}t_1 x_1+2\sqrt{\dac{\beta_2}{D}}t_2 x_2-t_1^2-t_2^2\bigg\}\dac{1}{D}\\
\times\exp\bigg\{-\dac{1}{2D}(\beta_1 x_1^2+\beta_2 x_2^2\bigg\}=\sum_{n_1=0}^{\infty}\sum_{n_2=0}^{\infty}\psi_{(n_1,n_2)}(\vec{x})\dac{t_1^{n_1}t_2^{n_2}}{n_1!n_2!}.
\end{gathered}
\label{exam2}
\end{equation}
Since the function $\phi(\vec{x})$ \eqref{exam1} has the symmetry $x_1 \leftrightarrow x_2$, we put
\begin{equation}
\beta_1=\beta_2=\beta.
\label{exam3}
\end{equation}
In view of \eqref{suppos3} and \eqref{exam3}, the~coefficient $k_n$ in \eqref{suppos3} is given by
\begin{equation}
\dil_{\BR^2}\varphi(\vec{x})h(\vec{x},t_1,t_2)d\vec{x}=\dac{\pi}{D\beta}\sum_{n_1=0}^{\infty}\sum_{n_2=0}^{\infty}k_n 2^{|n|} t_1^{n_1}t_2^{n_2}.
\label{exam4}
\end{equation}

The integral in \eqref{exam4} yields
\begin{equation}
\begin{gathered}
\dil_{\BR^2}\varphi(\vec{x})h(\vec{x},t_1,t_2)d\vec{x}=\dac{N}{D}
\bigg(\dac{2\pi\gamma_1}{1+\beta\gamma_1}\exp\bigg[\dac{\beta\gamma_1-1}{1+\beta\gamma_1}(t_1^2+t_2^2)\bigg]\\
-\dac{2\pi\varepsilon\gamma_2}{1+\beta\gamma_2}\exp\bigg[\dac{\beta\gamma_2-1}{1+\beta\gamma_2}(t_1^2+t_2^2)\bigg]\bigg),\\\\
k_{2n}=\dac{N\beta}{2^{|2n|}n!}\bigg(\dac{2\gamma_1}{1+\beta \gamma_1}\Big(\dac{\beta\gamma_1-1}{1+\beta\gamma_1}\Big)^{|n|}-\dac{2\varepsilon\gamma_2}{1+\beta \gamma_2}\Big(\dac{\beta\gamma_2-1}{1+\beta\gamma_2}\Big)^{|n|}\bigg), \qquad 2n=(2n_1,2n_2).
\end{gathered}
\label{exam5}
\end{equation}
Let the value of the parameter $\beta$ be considered optimal when the coefficients $k_{2n}$ converge to zero, as~$|n|\to \infty$ is the most rapid. It corresponds to the minimum value of the following~function:
\begin{equation}
f(\beta)=\max\bigg(\bigg|\dac{\beta\gamma_1-1}{1+\beta\gamma_1}\bigg|;\bigg|\dac{\beta\gamma_2-1}{1+\beta\gamma_2}\bigg|\bigg).
\label{exam6}
\end{equation}
The function $f(\beta)$ reaches its minimum at
\begin{equation}
\beta=\dac{1}{\sqrt{\gamma_1\gamma_2}}.
\label{exam7}
\end{equation}
We use this value of $\beta$ hereinafter. Note that the search of the value of $\beta_i$ corresponding to the the most rapid convergence can be more complicated in the general case. However, it is not crucial to obtain the exact value of $\beta_i$ corresponding to the least $|k_n|$ for large $|n|$. Hence, it can be done using some approximations for $k_n$ (e.g., \protect{\cite{gluzman2020}}).

The initial conditions for the EE system \eqref{eesyst2} can be obtained either by the substitution of the coefficients \eqref{exam5} into the formulae \eqref{suppos5} or by the substitution of the initial condition~\eqref{exam1} into the relations \eqref{incond2}. Both ways yield
\begin{equation}
\begin{gathered}
\sigma(0)=2N\pi(\gamma_1-\varepsilon \gamma_2), \\
\alpha_{11}^{(2)}(0)=\alpha_{22}^{(2)}(0)=D\dac{\gamma_1^2-\varepsilon\gamma_2^2}{\gamma_1-\varepsilon\gamma_2}.
\end{gathered}
\label{exam8}
\end{equation}

\begin{figure}[h]
\includegraphics[width=10.5 cm]{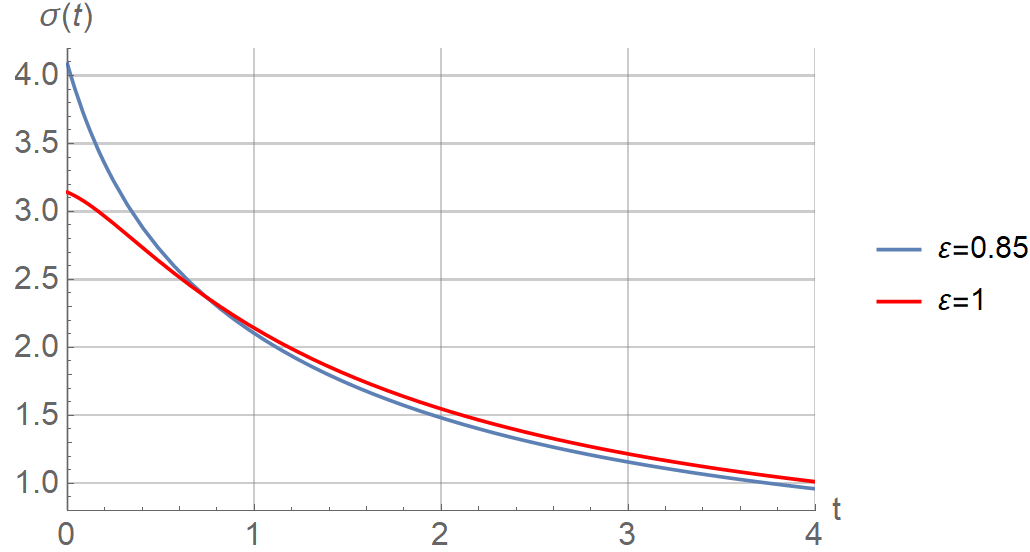}
\caption{The plot of the function $\sigma(t)$ for various $\varepsilon$. It illustrates the relaxation of the total number of ions for two initial distributions according to the analytic formula derived in
 \protect{\cite{shapkul21}} and {\eqref{exam8}}.\label{fig1}}
\end{figure}

\begin{figure}[h]
\includegraphics[width=10.5 cm]{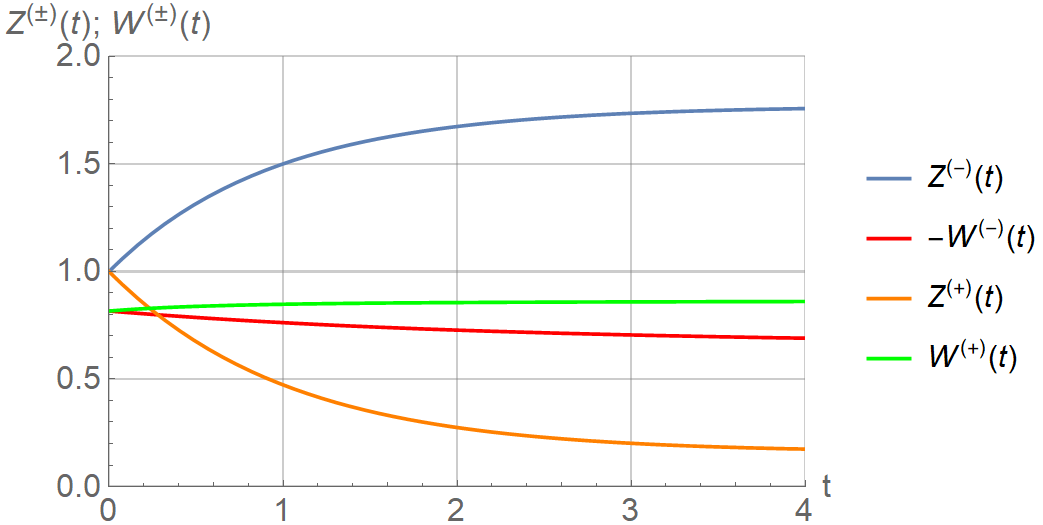}
\caption{The plot of the solutions to \eqref{var3} and \eqref{varsol2} for $\varepsilon=0.85$. The solutions are obtained~numerically.\label{fig2}}
\end{figure}

\begin{figure}[h]
\begin{minipage}[b][][b]{0.8\linewidth}\centering
    \includegraphics[width=10.5 cm]{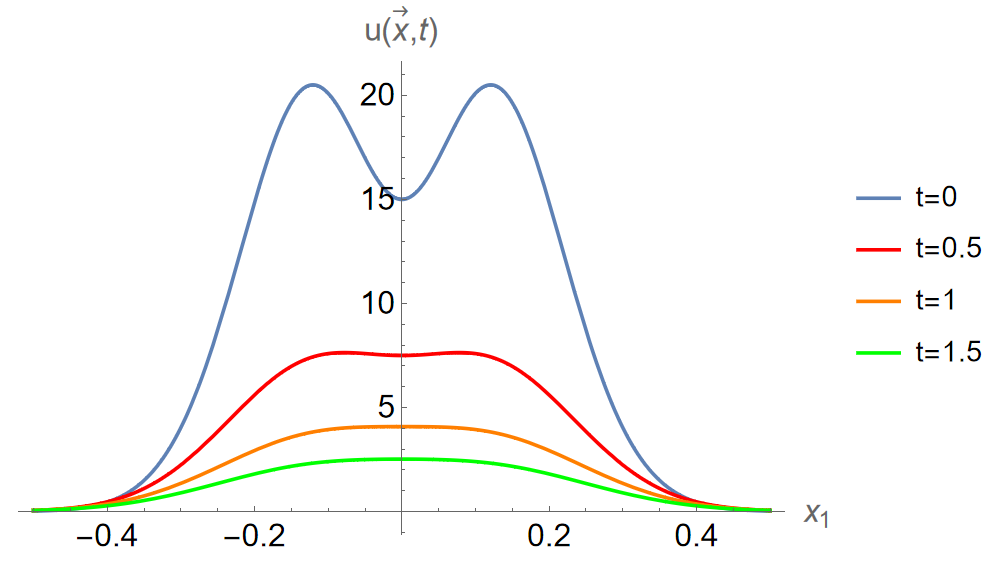} \\ a) $\varepsilon=0.85$
  \end{minipage}\\
  \begin{minipage}[b][][b]{0.8\linewidth}\centering
    \includegraphics[width=10.5 cm]{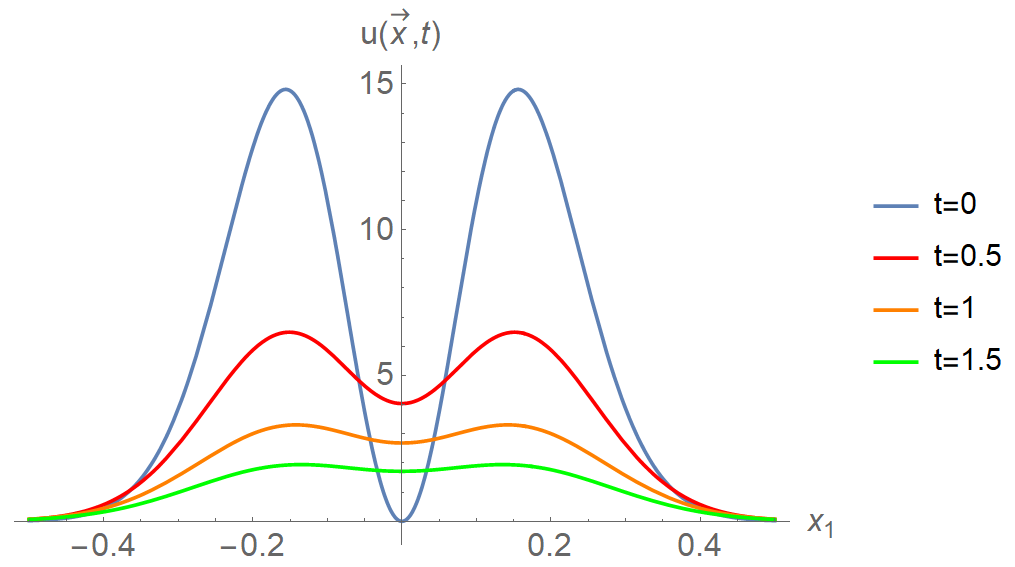} \\ b) $\varepsilon=1$
  \end{minipage}
\caption{The plot of the function $u(\vec{x},t)$ in the section $x_2=0$. It illustrates the evolution of the ion distribution according to the formula in {\eqref{suppos4}}.\label{fig3}}
\end{figure}

The EE system \eqref{eesyst2} is integrable for the coefficients in Equation \eqref{kineq1} of the form
\begin{equation}
a(t)=A_1 e^{-t/\tau_a}, \qquad \tilde{D}_a(t)=d_1 e^{-t/\tau_d}, \qquad b(t)=B_2+(B_1-B_2)e^{-t/\tau_b}.
\label{exam9}
\end{equation}
The general solutions to the EE system for the coefficients in \eqref{exam9} were obtained in~\cite{shapkul21}. This case is treated as the model of the plasma relaxation. Let us illustrate the solutions corresponding to the initial condition \eqref{exam1} for this case. Note that the variational system~\eqref{var3} is not integrable for such coefficients as far as we know. It is a remarkable fact, since the analytical solutions of the EE system, which are quite cumbersome though, can be expressed in terms of the solutions of the variational system by analogy with \eqref{genfun6}. For~our example, we use the solution to the EE system from~\cite{shapkul21} and construct the solutions to the variational system \eqref{var3} numerically. Note that the variational system \eqref{var3} is the system of linear differential equations with constant-sign coefficients. The~subsequent calculations are presented for $\varkappa=1$, $A_1=1$, $\tau_a=1$, $d_1=0.5$, $\tau_d=1$, $B_2=0.4$, $B_1=0.2$, $\tau_b=1$, $\mu=0.5$, $\gamma_1=1.5$, $\gamma_2=1$, $D=0.01$, $N=1$. Figure~\ref{fig1} shows the evolution of $\sigma(t)$ for two values of $\varepsilon$. The~physical meaning of this function is the total number of ions in the active medium. Figure~\ref{fig2} shows the solutions of the variational system. We provide the plot of these solutions just for one value of $\varepsilon$, since the difference between these plots for $\varepsilon=0.85$ and $\varepsilon=1$ is barely perceptible due to the little difference in $\sigma(t)$. The~sign of $W^{(-)}(t)$ was reverted for compactness of the figure. The~values of the expansion coefficients are presented in Table~\ref{tab1}. Note that they tend to zero rapidly as $|n|$ increases due to the optimal choice of $\beta$ and they equal zero for odd $n_1$ or $n_2$. Figure~\ref{fig3} shows the asymptotic evolution $u(\vec{x},t)$ of the initial ion distribution \eqref{exam1} constructed as the proposed nonlinear superposition in the weak diffusion approximation with an accuracy of $\Or(D^{3/2})$.

\unskip

\unskip


\begin{table}[h]
\caption{The values of the coefficient $k_{(n_1,n_2)}$ for $n_1,n_2=0,1,2,3,4$ and two values of $\varepsilon$.\label{tab1}}
		\begin{tabular}{|c|c|c|c|c|c|c|c|c|c|c|}
\hline
&  \multicolumn{5}{c|}{$\varepsilon=0.85$}  & \multicolumn{5}{c|}{$\varepsilon=1$}\\ \hline
\diagbox{$n_1$}{$n_2$} & 0 & 1 & 2 & 3 & 4 & 0 & 1 & 2 & 3 & 4 \\
\hline
  0 & $3.37\cdot 10^{-1}$ & 0 & $4.71\cdot 10^{-2}$ & 0 & $1.07\cdot 10^{-4}$ & $2.02\cdot 10^{-1}$ & 0 & $5.05\cdot 10^{-2}$ & 0 & $6.44\cdot 10^{-4}$ \\ \hline
  1 & 0 & 0 & 0 & 0 & 0 & 0 & 0 & 0 & 0 & 0 \\ \hline
  2 & $4.71\cdot 10^{-2}$ & 0 & $2.15\cdot 10^{-4}$ & 0 & $1.50\cdot 10^{-5}$ & $5.05\cdot 10^{-2}$ & 0 & $1.29\cdot 10^{-4}$ & 0 & $1.61\cdot 10^{-5}$ \\ \hline
  3 & 0 & 0 & 0 & 0 & 0 & 0 & 0 & 0 & 0 & 0 \\ \hline
  4 & $1.07\cdot 10^{-4}$ & 0 & $1.50\cdot 10^{-5}$ & 0 & $3.43\cdot 10^{-8}$ & $6.44\cdot 10^{-5}$ & 0 & $1.61\cdot 10^{-5}$ & 0 & $2.05\cdot 10^{-8}$ \\
\hline
\end{tabular}
\end{table}

Note that $Z^{(+)}(t)$ does not tend to nonzero constant. In~a general case, this constant can be negative, which leads to the zero value of $Z^{(+)}(t)$ at some point $t=t_0$. At~this point, the~condition of nondegeneracy of the matrix $C(t)$ \eqref{ric2} is violated. It is worthy of discussion how it affects the solutions $v_n(\vec{x},t,\BFC_n)$. It can be shown that the functions \eqref{genfun6} have the removable discontinuity at this point. The~same is true for the solutions in \eqref{exc2}. Hence, the~asymptotic solutions $u(\vec{x},t)=v_n(\vec{x},t,\BFC_n)$ regularly depend on $t$ in a neighborhood of the point $t=t_0$. It means that if we construct the germ $r^{2}_t$ from the point $t=0$ to a point left of the point $t=t_0$ and then we construct the germ $\tilde{r}^{2}_t$ after the point $t=t_0$, the~resulting asymptotic solutions $u(\vec{x},t)$, $t<t_0$, and~$\tilde{u}(\vec{x},t)$, $t>t_0$, generated by the germ $r^{2}_t$ and $\tilde{r}^{2}_t$, respectively, determine the continuous asymptotic evolution of the initial state $u(\vec{x},t)\big|_{t=0}$ for both $t<t_0$ and $t>t_0$. Therefore, the~formal constraint of the nondegeneracy of the matrix $C(t)$ \eqref{ric2} can be bypassed. Note that the error of the semiclassical solutions usually grows over time. It can result in an adverse effect when the semiclassical solution jumps from the asymptotics for one exact solution to another at a sufficiently large time (see the work~\cite{bagrov98}). Therefore, the~estimate of the period of time where the asymptotics are valid is the subject to study separately and is to be obtained in every particular~case.

Figure~\ref{fig1} illustrates that the ion distribution with the less pronounced minimum at the GDT center relaxes faster. However, Figure~\ref{fig3} shows that both distributions tend to the Gauss-like profile over time. It means that their evolution will be similar on large times. Hence, the~difference is significant only at the initial stage of the~relaxation.

Note that our method is applicable for the linear case ($\varkappa=0$). For~$\varkappa=0$ and $a(\vec{x},t)=a(t)$, the~method proposed yields the exact solution to the kinetic equation.

\section{Conclusions}

\label{concl}
Within the framework of the approach presented in~\cite{shapkul21}, we have constructed a countable family of asymptotic solutions to the two-dimensional kinetic Equation \eqref{kineq1} with the nonlocal cubic nonlinearity. The~approach proposed can be generalized for $n$-dimensional space. We have considered the two-dimensional problem for the sake of simplicity in order to demonstrate our method and to give its physical interpretation. The~constructed asymptotic solutions correspond to the weak diffusion approximation. The~approach is based on the solutions to the Cauchy problem to the nonlinear dynamical system (the Einstein--Ehrenfest system) \eqref{eesyst1} and \eqref{eesyst2}. These solutions generate the linear parabolic \mbox{Equations \eqref{ale2} and \eqref{ale4}} associated with the nonlinear kinetic equation. Solutions of this associated linear equation subject to the algebraic condition \eqref{alcond1} yield the asymptotic solutions \eqref{exc2} and \eqref{cons1} to the original nonlinear kinetic equation. The~family of the asymptotic solutions is constructed based on the set of skew-orthogonal, linearly independent solutions to the system of linear ODEs \eqref{var3} (the variational system). Such asymptotic solutions form the orthogonal basis for the nonlinear superposition principle that allows one to construct the asymptotic evolution of arbitrary initial distribution $\varphi(\vec{x})$ \eqref{incond3} expanded into the series with respect to the basis formed by the functions \eqref{exc2}. The~nonlinearity of the superposition principle is caused by the necessity of solving the nonlinear dynamical system \eqref{eesyst1} and \eqref{eesyst2} with initial conditions \eqref{incond2} determined by the initial distribution. Note that the functions $v_n(\vec{x},t,\BFC)$ with $\BFC$ determined from {\eqref{suppos5}} form an orthogonal basis for the nonlinear superposition principle, while the functions $v_n(\vec{x},t,\BFC_n)$ with $\BFC_n$ given by {\eqref{cons3}} and {\eqref{cons4}}, which determine independent asymptotic solutions to the nonlinear kinetic equation, are not orthogonal in the general case.

This work extends the results of our work~\cite{shapkul21} where the asymptotic evolution operator of the kinetic Equation \eqref{kineq1} was constructed via the Green function of the associated linear equation. Here, we construct the asymptotic solutions with the help of the symmetry operators \eqref{sym8} for the associated linear equation based on the nonlinear superposition principle. It allows us to study the properties of the asymptotic solutions through the properties of the variational system \eqref{var3} that generates the Maslov germ $r^2_t$ on the zero-dimensional manifold $\Lambda^0_t$ and the symmetry operators. Moreover, we have found the relation \eqref{genfun6} between the solutions to the Einstein--Ehrenfest system and the variational system. It means that our approach can yield analytical solutions expressed in terms of the solutions to the system of the linear ODEs \eqref{var3} only. However, in~the considered specific case, the~nonlinear dynamical system admits the analytical solutions, while the variational system \eqref{var3} is solved numerically. Note that the asymptotic evolution operator in~\cite{shapkul21} and the nonlinear superposition principle yield the exact solutions to the associated linear Equation \eqref{ale4}, i.e.,~they yield the same asymptotic solutions for the given initial condition. However, the~approach proposed here broadens the possibilities of analysis for these solutions. In~addition, the~nonlinear superposition principle leads to the expressions with free parameters that can be chosen so that the functional series \eqref{suppos4} converges rapidly. Hence, in~some cases, the~study of the solutions \eqref{suppos4} can be practically reduced to the study of the first few terms of the series. It makes the analysis of such solutions even easier, especially when the solutions are~numerical--analytical.

The future prospects for this work are related to the study of a more complex two-component kinetic model. We plan to generalize the approach~\cite{fkppshap19} for the latter problem.

\section*{Acknowledgement}
The reported study was funded by RFBR and Tomsk region according to the research project No. 19-41-700004. The work is supported by Tomsk Polytechnic University under the International Competitiveness Improvement Program; by IAO SB RAS, Russia, project no. 121040200025-7.

\bibliography{lit1}

\end{document}